\documentclass[letterpaper]{article} 
\usepackage{aaai2027}  
\usepackage[hyphens]{url}  
\usepackage{graphicx} 
\usepackage{natbib}  
\usepackage{caption} 
\usepackage{algorithm}
\usepackage{algorithmic}
\usepackage{amsmath} 
\usepackage{amssymb}
\usepackage{multirow}
\usepackage{newfloat}
\usepackage{listings}
\DeclareCaptionStyle{ruled}{labelfont=normalfont,labelsep=colon,strut=off} 
\floatstyle{ruled}
\newfloat{listing}{tb}{lst}{}
\floatname{listing}{Listing}

\usepackage{booktabs}

\title{REIMU: Efficient Heterogeneous Hierarchical Reasoning for SSL-Based Speech Deepfake Detection}
\author{
}
\affiliations{
    \textsuperscript{\rm 1}College of Cyber Security, Jinan University, Guangzhou, China\\

    proceedings-questions@aaai.org
}

\title{REIMU: Efficient Heterogeneous Hierarchical Reasoning for SSL-Based Speech Deepfake Detection}
\author {
    Kwok-Ho Ng\textsuperscript{\rm 1,}
    Tingting Song\textsuperscript{\rm 1}\corresponding,
    Bingwen Feng\textsuperscript{\rm 1},
    Peiya Li\textsuperscript{\rm 1}
}
\affiliations {
    \textsuperscript{\rm 1} College of Cyber Security, Jinan University, Guangzhou, China\\
    kwokhong@stu2024.jnu.edu.cn, tingtingsong@jnu.edu.cn
}

\begin{document}
\nocopyright
\maketitle

\begin{abstract}
The increasing realism of speech generated by text-to-speech and voice conversion systems poses growing challenges to media integrity and voice authentication. Self-supervised learning (SSL) has substantially advanced speech deepfake detection, where downstream backbones conventionally process SSL representations through a single forward pass. This work investigates the practical effectiveness of recurrent hierarchical reasoning for this task. We term this controlled study REIMU and systematically compare conventional single-pass backbones, weight-shared recurrence, homogeneous HRM, and heterogeneous HRM across four Base-scale SSL frontends. We further examine heterogeneous high- and low-level modules that combine self-attention with linear attention. Experiments on the ASVspoof 2019 and 2021 evaluation sets show that recurrence and hierarchical decomposition do not inherently improve detection, whereas heterogeneous operator assignment provides a more competitive configuration. Notably, the heterogeneous design remains competitive while using 10.8\% fewer downstream parameters than the matched baseline, demonstrating its potential for parameter-efficient speech deepfake detection.

\end{abstract}

\begin{links}
    \link{Code}{https://github.com/saki-ciallo/REIMU-SDD}
\end{links}

\section{Introduction}
Recent advances in text-to-speech (TTS) synthesis and voice conversion (VC) have enabled high-quality synthetic speech to be widely adopted in content creation and assistive applications \cite{hu2026qwen3}. However, the malicious deployment of these technologies facilitates voice cloning, identity impersonation, misinformation, and presentation attacks against voice authentication systems, posing concrete threats to digital media integrity and information security \cite{li2025survey}. To counter these risks, benchmarks such as the ASVspoof series \cite{wang2020asvspoof,yamagishi2021asvspoof} have continuously driven research on speech anti-spoofing and deepfake detection. The AT-ADD challenge \cite{xie2026add} further broadens the scope to realistic conditions across diverse audio types, including speech, environmental sounds, singing, and music, while multilingual datasets like MLAAD \cite{muller2024mlaad} evaluate generalization to unseen languages and synthesis engines. Consequently, exploring and constructing reliable speech deepfake detection (SDD) systems has become a critical research priority in audio security.

Early SDD architectures focused on learning discriminative spoofing cues directly from raw waveforms. For instance, RawNet2 established end-to-end neural network paradigms to learn artifact patterns without relying on handcrafted acoustic features \cite{tak2021end}. Building upon raw waveform modeling, RawGAT-ST \cite{tak2021end} introduced joint temporal and spectral graph attention to capture spoofing artifacts across time and frequency regions effectively. Subsequently, AASIST \cite{jung2022aasist} incorporated heterogeneous spectro-temporal graph attention and graph pooling strategies, becoming a widely adopted end-to-end baseline architecture in speech anti-spoofing research.

With the rapid progress of self-supervised speech learning, encoders pretrained on massive unlabelled corpora can be effectively transferred to SDD. Combining an SSL frontend with a downstream classifier has thus emerged as a dominant paradigm. For instance, XLSR-AASIST \cite{tak2022automatic} utilized wav2vec 2.0 XLSR (300M) as a feature extractor. It achieved substantial performance gains on the ASVspoof 2021 evaluation sets. Similar efforts introduced Conformer backbones combining convolution and self-attention \cite{rosello2023conformer}. Additionally, XLSR-Mamba \cite{xiao2025xlsr} leveraged a dual-column bidirectional Mamba backbone to model local and long-range dependencies with linear complexity. Beyond backbone enhancements, other research explores extracting richer representations from SSL encoders \cite{el2025comprehensive}. For example, XLSR-SLS \cite{zhang2024audio} treated all intermediate layers of the pretrained Transformer as hierarchical features and learned sensitivity weights for aggregation. This approach leveraged the rich spoofing artifacts preserved across different SSL hidden layers. Similarly, parameter-efficient adaptation techniques demonstrate effective representation alignment with minimal trainable parameters. A prominent example learns wavelet prompts inside a frozen SSL encoder paired with an AASIST backbone \cite{xie2026detect}.

Overall, existing studies primarily enhance modeling capacity by adapting SSL frontends, aggregating layer-wise features, or designing complex downstream backbones. Nevertheless, most backbones still transform input features through a single forward pass of fixed depth. It remains insufficiently explored whether a compact backbone can iteratively refine latent representations through parameter reuse rather than continuously adding new parameters. Specifically, when the SSL frontend already yields frame-level features, the impact of weight-shared recurrence on artifact extraction is not fully understood. Furthermore, it is worth examining whether modules with different update frequencies should employ complementary sequence-modeling operators.

To achieve deep computation within constrained parameter budgets, the Hierarchical Reasoning Model (HRM) \cite{wang2025hierarchical} was recently proposed. It utilizes two interacting recurrent modules operating at distinct update frequencies. The high-level module updates an abstract latent state at a lower frequency, while the low-level module performs granular computations more frequently. These intermediate operations are conducted entirely in continuous latent space. The subsequent HRM-Text framework \cite{wang2026hrm} extended this architecture to language modeling. It achieved competitive performance against larger open models while using substantially less pretraining data and compute.

However, the underlying mechanisms of the dual-module H/L architecture remain actively discussed in the literature. Subsequent studies reported that simple, approximately equivalent Transformer \cite{vaswani2017attention} baselines can achieve comparable performance under equivalent settings \cite{ge2025hierarchical}. These observations suggest that performance gains from recurrent computation should not be automatically attributed to hierarchical design alone. Therefore, when transferring HRM to speech deepfake detection, it is essential to reframe it as a mechanism for recurrent latent feature refinement. Furthermore, it is necessary to systematically isolate the distinct effects of standard single forward pass depth, simple weight-shared recurrence, and true hierarchical recurrence with separately parameterized modules.

Motivated by these considerations, this paper investigates the applicability of recurrent hierarchical architectures for speech deepfake detection using self-supervised representations. We refer to this controlled investigation of recurrent and heterogeneous hierarchical processing for SSL-based SDD as REIMU. Unlike symbolic reasoning, SDD requires identifying subtle local acoustic artifacts from continuous frame-level features and integrating evidence across an utterance. We hypothesize that the H and L modules do not necessarily benefit from identical sequence operators. Specifically, the low-level module can employ linear attention (e.g., Gated DeltaNet-2 \cite{hatamizadeh2026gated} or Raven \cite{afzalbick2026raven}) for low-cost recurrent refinements on local features. Meanwhile, the high-level module can use multi-head self-attention (MHSA) for global temporal modeling over refined states.

To validate this hypothesis, we establish a controlled benchmarking framework using four 95M-parameter (approximately) SSL frontends: wav2vec 2.0 Base \cite{baevski2020wav2vec}, HuBERT Base \cite{hsu2021hubert}, WavLM Base \cite{chen2022wavlm}, and WavLM Base+. Under frozen-encoder settings, we systematically evaluate standard single forward pass baselines, simple weight-shared Looped models \cite{yang2024looped}, and both homogeneous and heterogeneous HRM configurations across multiple execution schedules ($H_2L_1, H_2L_2, H_2L_3$). In Looped models, gradients participate only in the final module call. Based on overall performance, Gated DeltaNet-2 (GDN2) is selected as the core sequence operator for further ablation studies. We unfreeze top SSL layers and apply a unified data augmentation setup to assess the performance gains of selective adaptation.

The primary contributions of this work are summarized as follows:
\begin{itemize}
\item To the best of our knowledge, this work innovatively presents a controlled investigation of hierarchical recurrent architectures for SDD.

\item We conduct a unified comparison across four 95M-parameter SSL frontends under frozen-encoder conditions, assessing downstream architectural performance and generalization across diverse speech representations.

\item We explicitly disentangle the effects of standard single forward pass computation, single-module weight-shared Looped refinement, homogeneous HRM, and heterogeneous HRM under truncated-gradient constraints.

\item We investigate heterogeneous operator assignment across update frequencies by combining a high-level MHSA module with low-level linear attention operators (GDN2 and Raven). Experimental results demonstrate that under matched settings, introducing operator heterogeneity (e.g., $H_{\mathrm{MHSA}} + L_{\mathrm{GDN2}}$) reduces downstream backbone parameters by 10.8\% while achieving competitive equal error rates relative to standard baselines on the ASVspoof 2019 and 2021 evaluation sets.
\end{itemize}

\section{Methodology}

\subsection{Framework Definition}

\begin{figure*}[t]
\centering
\includegraphics[width=0.95\textwidth]{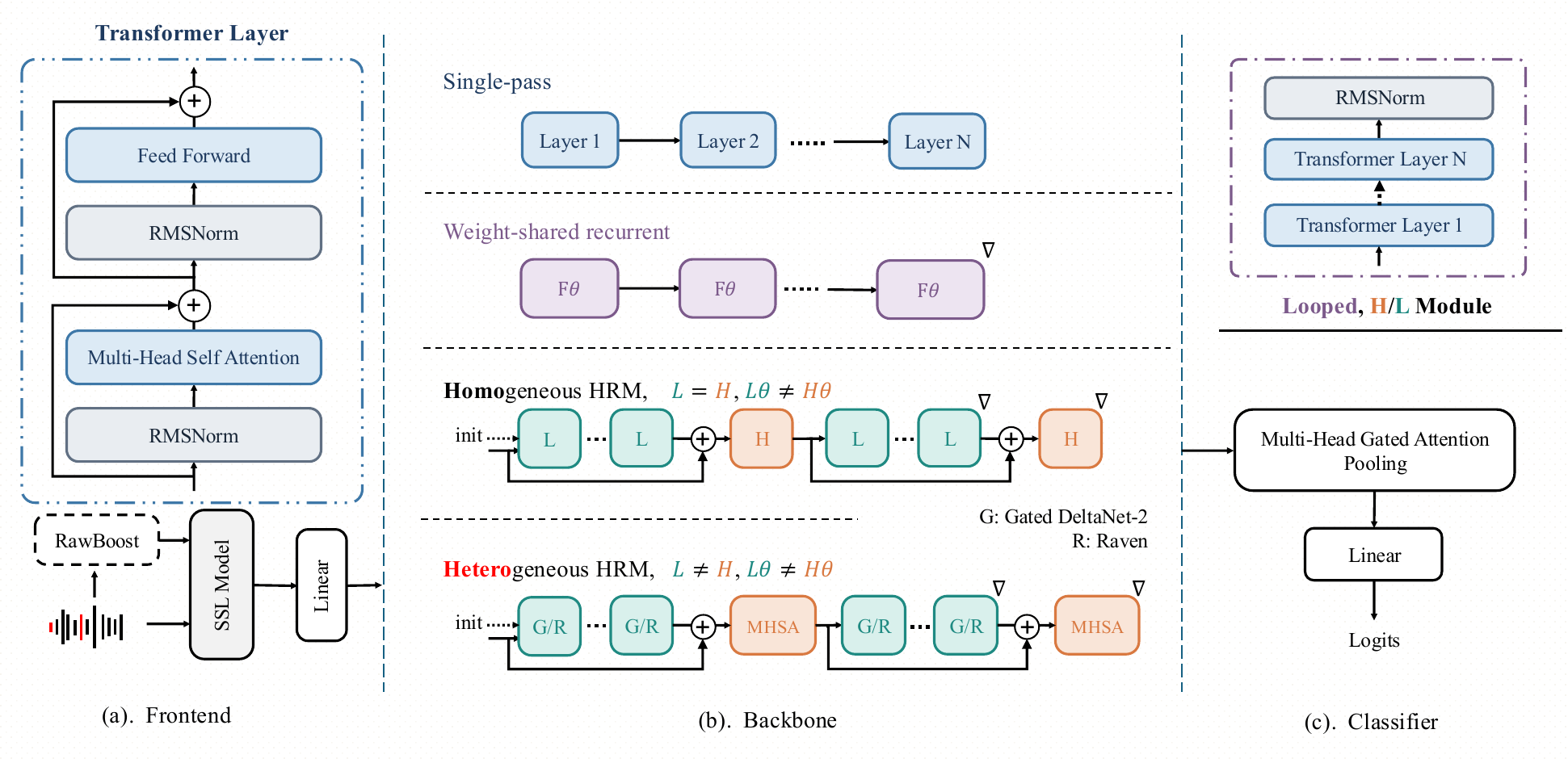} 
\caption{Overview of the structure of the designed experiment.}
\label{fig:reimu}
\end{figure*}

Figure \ref{fig:reimu} illustrates the pipeline of the proposed experimental method. Given a raw speech waveform $x \in \mathbb{R}^T$ of length $T$, the goal of SDD is to predict a continuous spoofing score $\hat{y} \in [0, 1]$, where values closer to $0$ and $1$ represent bonafide and spoofed speech, respectively. 
\begin{equation}
    \hat{y} = C\big(P(B(\mathrm{Linear}(F(x)))\big) \label{eq:df}
\end{equation}
where $F(\cdot)$ denotes the frozen or partially fine-tuned SSL frontend, $\mathrm{Linear}(\cdot)$ maps the SSL representations to the hidden dimension of the backbone, $B(\cdot)$ represents the sequence modeling backbone, $P(\cdot)$ denotes the pooling operator, and $C(\cdot)$ is the final binary classifier.


\subsection{SSL Frontend}
To evaluate the impact of frontend representations on downstream backbones under controlled conditions, we select four SSL encoders with comparable parameter scale ($\approx 95$M) and identical output embedding dimensions. Given an input waveform $x$, the feature extraction process is expressed as:

\begin{equation}
    z_{\mathrm{ssl}} = F(x) \in \mathbb{R}^{S \times D}, \label{eq:ssl}
\end{equation}
where $S$ denotes the subsampled frame sequence length and $D$ represents the SSL hidden dimension.

To bridge the dimension gap between various SSL frontends and downstream backbones, a linear projection layer projects $z_{\mathrm{ssl}}$ to the backbone model dimension $d$:
\begin{equation}
    z_{0} = \mathrm{Linear}(z_{\mathrm{ssl}}) \in \mathbb{R}^{S \times d}. \label{eq:ssl2backbone}
\end{equation}

In main architectural comparisons, all parameters of $F(\cdot)$ remain strictly frozen to isolate the downstream architectural effects. In subsequent fine-tuning experiments, we unfreeze only the last two Transformer blocks of the SSL encoder for selective adaptation.

\subsection{Unified Backbone and Sequence Mixers}
To ensure fair comparisons across different sequence operators, we construct a unified Transformer-like backbone. The backbone consists of $N$ stacked blocks, where each block adopts a pre-normalization (RMSNorm \cite{zhang2019root}) structure with a sequence mixing module and a SwiGLU \cite{shazeer2020glu} Feed-Forward Network (FFN):
\begin{align}
    z^{(l) \prime} &= z^{(l-1)} + M\Big(\mathrm{RMSNorm}(z^{(l-1)})\Big), \\ 
    z^{(l)} &= z^{(l) \prime} + \mathrm{SwiGLU}\Big(\mathrm{RMSNorm}(z^{(l) \prime})\Big),
\end{align}
where $l \in \{1, \dots, N\}$ indicates the block index, and $M(\cdot)$ represents the candidate sequence mixing operator. All candidate operators share identical hyperparameters, including the number of attention/state heads and hidden projections.

We evaluate three sequence mixing operators with distinct properties:
\begin{itemize}
    \item \textbf{MHSA} captures global temporal dependencies through pairwise content interactions. It is used in the Transformer baselines and as the high-level operator in heterogeneous HRM.

    \item \textbf{GDN2} compresses temporal history into a fixed-size matrix state with gated linear-complexity updates, making it suitable for frequent low-level refinement.
    
    \item \textbf{Raven} maintains routed memory slots that are selectively updated to capture localized and transient acoustic artifacts.
    
\end{itemize}

\subsubsection{Standard Single Forward Pass}
The standard single forward pass baseline serves as the conventional feed-forward reference. It consists of $N$ sequentially stacked backbone blocks with independent parameters:
\begin{equation}
    z^{(l)} = B_l\left(z^{(l-1)}\right), \quad l \in \{1, 2, \dots, N\}, \label{eq:standard}
\end{equation}
where $z^{(0)} = z_0$ is the linearly projected SSL representation from Eq.~(\ref{eq:ssl2backbone}), and $B_l(\cdot)$ denotes the $l$-th distinct backbone block parameterized by an independent parameter set $\Theta_l$. The final output $z^{(N)}$ is passed to the pooling layer.

To establish fair control baselines, we fix the network depth $N$  across all standard architectures and strictly alter only the internal sequence mixing operator $M(\cdot)$ inside $B_l$. By substituting this operator, we construct three feed-forward baselines: \textbf{MHSA-FFN}, \textbf{GDN2-FFN}, and \textbf{Raven-FFN}. These baselines represent conventional Transformer-style architectures where feature transformations are executed through a single pass over independently parameterized layers.

\subsubsection{Weight-Shared Looped Refinement}
To isolate the effects of iterative computation from hierarchical state decomposition, we construct a parameter-shared recurrent baseline. Rather than stacking $N$ unique blocks, a single composite backbone module $B_{\mathrm{shared}}$ with parameter set $\Theta_{\mathrm{shared}}$ is applied repeatedly over $R$ recurrent steps in continuous latent space.

To maintain numerical stability during multi-turn recurrent updates and use a similar design in HRM, the recurrent module is structured with an outer post-normalization layer:
\begin{equation}
    B_{\mathrm{shared}}(x) = \mathrm{RMSNorm}\Big(\mathrm{Transformer}(x)\Big), \label{eq:postnorm_module}
\end{equation}
where $\mathrm{Transformer}(\cdot)$ adopts the unified pre-norm block. The composite block $B_{\mathrm{shared}}$ is then treated as a unified atomic unit for recurrent execution:
\begin{equation}
    z^{(r)} = B_{\mathrm{shared}}\left(z^{(r-1)}\right), \quad r \in \{1, 2, \dots, R\}, \label{eq:looped_forward}
\end{equation}
where $z^{(0)} = z_0$, and $R$ specifies the total number of recurrent passes. All recurrent steps share the exact same parameters $\Theta_{\mathrm{shared}}$.

To ensure memory efficiency during training and prevent gradient explosion, backpropagation is applied exclusively to the final recurrent iteration. Formally, let $\text{sg}(\cdot)$ denote the stop-gradient operator. The complete execution flow is expressed as:
\begin{align}
    \tilde{z}^{(0)} &= z_0, \nonumber \\
    \tilde{z}^{(r)} &= \text{sg}\left(B_{\mathrm{shared}}\left(\tilde{z}^{(r-1)}\right)\right), \quad \text{for } r = 1, \dots, R-1, \label{eq:looped_sg} \\
    z^{(R)} &= B_{\mathrm{shared}}\left(\tilde{z}^{(R-1)}\right). \label{eq:looped_final}
\end{align}
Under this formulation, gradients flow strictly through the final iteration step $R$, while the previous $R-1$ iterations serve solely to refine the latent representation in a forward-only manner. This baseline enables us to examine whether simple parameter-shared recurrence over a single composite module can match the refinement efficacy of hierarchical H/L architectures.

\subsubsection{Homogeneous HRM}
To achieve deeper effective computation without expanding parameter budgets, the Hierarchical Reasoning Model (HRM) decomposes the backbone into two interacting modules: a High-level module $H(\cdot; \Theta_H)$ and a Low-level module $L(\cdot; \Theta_L)$. To strictly control total parameter capacity, both $H$ and $L$ modules are constructed with $N/2$ Transformer-style blocks (half the depth of the $N$-block standard baseline). Thus, the combined parameter count of $H$ and $L$ matches a single standard $N$-block baseline ($\Theta_H + \Theta_L \approx \Theta_{\mathrm{standard}}$). In Homogeneous HRM, $H$ and $L$ employ identical sequence mixing operator types (e.g., both using MHSA or GDN2).

The execution follows a hierarchical schedule denoted as $H_2L_k$ ($k \in \{1, 2, 3\}$), where $2$ specifies the total number of outer High-level cycles, and $k$ dictates the number of inner Low-level recurrent refinements preceding each High-level step.

Let $z_0$ denote the projected SSL representation from Eq.~(\ref{eq:ssl2backbone}), and $h_{\mathrm{init}} \in \mathbb{R}^{S \times d}$ denote a learnable initial high-level state. The recurrent state initialization is formulated as:
\begin{equation}
    h^{(0)} = h_{\mathrm{init}}, \quad z_{\mathrm{in}}^{(1, 1)} = z_0 + h^{(0)}, \label{eq:hrm_init}
\end{equation}
where $z_{\mathrm{in}}^{(1, 1)}$ serves as the input to the very first Low-level step.

Formally, for the $m$-th High-level cycle ($m \in \{1, 2\}$) and its $n$-th Low-level sub-step ($n \in \{1, \dots, k\}$), the state transitions and residual fusion mechanisms are defined recursively as:
\begin{align}
    \text{Low Input:} \quad z_{\mathrm{in}}^{(m, n)} &= 
    \begin{cases} 
        z_0 + h^{(m-1)}, & \text{if } n = 1, \\
        z_{\mathrm{out}}^{(m, n-1)}, & \text{if } n > 1,
    \end{cases} \label{eq:hrm_low_in} \\
    \text{Low-Level:} \quad z_{\mathrm{out}}^{(m, n)} &= L\left(z_{\mathrm{in}}^{(m, n)}; \Theta_L\right), \label{eq:hrm_low_exec} \\
    \text{High Input:} \quad h_{\mathrm{in}}^{(m)} &= h^{(m-1)} + z_{\mathrm{out}}^{(m, k)}, \label{eq:hrm_high_in} \\
    \text{High-Level:} \quad h^{(m)} &= H\left(h_{\mathrm{in}}^{(m)}; \Theta_H\right), \label{eq:hrm_high_exec}
\end{align}
where Eq.~(\ref{eq:hrm_high_in}) explicitly fuses the high-level latent state from the previous cycle $h^{(m-1)}$ with the output of the current final Low-level step $z_{\mathrm{out}}^{(m, k)}$ ($H_1\_output + L_2\_output$ for $m=2$) before feeding into the High-level module $H$.

Under the $H_2L_k$ schedule, the explicit execution sequences are unrolled as follows:
\begin{itemize}
    \item \textbf{$H_2L_1$ ($k=1$):} $L \rightarrow H \rightarrow L \rightarrow H$ (2 outer cycles, each preceded by 1 $L$-pass).
    \item \textbf{$H_2L_2$ ($k=2$):} $L \rightarrow L \rightarrow H \rightarrow L \rightarrow L \rightarrow H$ (2 outer cycles, each preceded by 2 $L$-passes).
    \item \textbf{$H_2L_3$ ($k=3$):} $L \rightarrow L \rightarrow L \rightarrow H \rightarrow L \rightarrow L \rightarrow L \rightarrow H$ (2 outer cycles, each preceded by 3 $L$-passes).
\end{itemize}

Similar to the Looped baseline, gradient propagation is truncated during intermediate recurrent steps. Backpropagation is executed strictly through the final module calls of the $H_2L_k$ sequence using the stop-gradient operator $\text{sg}(\cdot)$, preventing gradient instability across unrolled computation graphs.

\subsubsection{Heterogeneous HRM}
While Homogeneous HRM enforces $H$ and $L$ to share the same sequence mixing operator, we hypothesize that modules operating at different update frequencies require distinct sequence-modeling inductive biases. To investigate this structural hypothesis, we propose the Heterogeneous Hierarchical Architecture (Hetero-HRM).

Hetero-HRM retains the $N/2$-block $H$ and $N/2$-block $L$ setup, the residual fusion mechanism, and the $H_2L_k$ schedule, but decouples the choice of sequence operators across temporal frequencies:
\begin{itemize}
    \item \textbf{High-Level Module ($H_{\mathrm{MHSA}}$):} Operating at a lower execution frequency (2 cycles total), the high-level module consistently employs MHSA. It performs global pairwise temporal modeling over the composite features $h_{\mathrm{in}}^{(m)}$ to aggregate utterance-level spoofing evidence.
    \item \textbf{Low-Level Module ($L_{\mathrm{Linear}}$):} Operating at a high execution frequency ($k$ steps per cycle), the low-level module is instantiated with efficient linear attention operators—specifically testing GDN2 and Raven. These operators feature $O(S)$ sequence complexity and dynamic state/slot update mechanisms, enabling rapid, low-cost recurrent refinements of local acoustic artifact boundaries.
\end{itemize}
By systematically evaluating different linear sequence mixers within the low-level module ($L_{\mathrm{GDN2}}$ vs. $L_{\mathrm{Raven}}$) alongside a high-level MHSA module, Hetero-HRM investigates whether operator heterogeneity across update frequencies serves as a general and effective strategy for recurrent speech deepfake detection.

\subsection{Pooling and Classifier}

To aggregate frame-level features $Z \in \mathbb{R}^{S \times d_{\mathrm{in}}}$ into a fixed-dimensional utterance representation $z_{\mathrm{pooled}}$, we employ Multi-Head Gated Attention Pooling (MHGAP). MHGAP combines frame-wise channel gating with multi-head temporal attention to adaptively aggregate spoofing evidence.

Given the input sequence $Z$, a linear projection produces the value and gate features:
\begin{equation}
    [\mathbf{V}, \mathbf{G}] = \mathrm{Linear}(Z) \in \mathbb{R}^{S \times (2d_{\mathrm{pooled}})}.
    \label{eq:mhgap_proj}
\end{equation}
The projected features are divided into $K$ heads ($d_k = d_{\mathrm{pooled}} / K$), yielding $\mathbf{V}_{s,k}, \mathbf{G}_{s,k} \in \mathbb{R}^{d_k}$.

For frame $s \in \{1,\ldots,S\}$ and head $k \in \{1,\ldots,K\}$, the gated evidence vector is computed as:
\begin{equation}
    \mathbf{E}_{s,k} = \mathbf{V}_{s,k} \odot \mathrm{SiLU}(\mathbf{G}_{s,k}).
    \label{eq:mhgap_gate}
\end{equation}

In parallel, each head is associated with a learnable scoring vector $\mathbf{w}_k \in \mathbb{R}^{d_k}$. The normalized temporal attention weight $a_{s,k}$ is computed via a temperature-scaled Softmax:
\begin{equation}
    a_{s,k} = \frac{\exp\left( \mathbf{V}_{s,k}^{\top}\mathbf{w}_k / \tau \right)}{\sum_{s'=1}^{S} \exp\left( \mathbf{V}_{s',k}^{\top}\mathbf{w}_k / \tau \right)},
    \label{eq:mhgap_attn}
\end{equation}
where $\tau > 0$ denotes the score temperature.

The representation for head $k$ is obtained by aggregating the gated evidence over the temporal dimension:
\begin{equation}
    \mathbf{P}_k = \sum_{s=1}^{S} a_{s,k}\mathbf{E}_{s,k}.
    \label{eq:mhgap_head_agg}
\end{equation}

The head representations are concatenated and regularized by dropout:
\begin{equation}
    z_{\mathrm{pooled}} = \mathrm{Dropout}\left( \operatorname{Concat}\left( \mathbf{P}_1,\ldots,\mathbf{P}_K \right) \right).
    \label{eq:mhgap_out}
\end{equation}

Finally, a linear classifier maps the pooled representation to two class logits $\boldsymbol{\ell} = W_{\mathrm{cls}}z_{\mathrm{pooled}} \in \mathbb{R}^{2}$. The predicted class is obtained via:
\begin{equation}
    \hat{y} = \arg\max_{c \in \{0,1\}} \ell_c.
    \label{eq:classifier_prediction}
\end{equation}

\begin{table}[t]
    \centering
    \small
    \begin{tabular}{llrrr}
        \toprule
        Dataset & Partition & Bonafide & Spoof & Total \\
        \midrule
        19LA & Train        & 2,580  & 22,800  & 25,380 \\
        19LA & Dev.  & 2,548  & 22,296  & 24,844 \\
        \midrule
        19LA & Eval. & 7,355  & 63,882  & 71,237 \\
        21LA & Eval. & 14,816 & 133,360 & 148,176 \\
        21DF & Eval. & 14,869 & 519,059 & 533,928 \\
        \bottomrule
    \end{tabular}
    \caption{
        Number of bonafide and spoofed utterances in the datasets used for ASVspoof 2019/2021 training, development, and evaluation.
    }
    \label{tab:dataset_statistics}
\end{table}

\section{Experiment Setup}
\subsection{Datasets and Metrics}
All models are trained exclusively on the ASVspoof 2019 Logical Access (19LA) training set, with model selection performed on its development set. We evaluate all systems across three evaluation benchmarks: 19LA, ASVspoof 2021 Logical Access (21LA), and ASVspoof 2021 Deepfake (21DF). The 19LA evaluation set contains previously unseen spoofing attack configurations based on TTS and VC systems. The 21LA evaluation set assesses robustness to telephony and Voice over Internet Protocol (VoIP) transmission and codec distortions, whereas 21DF evaluates cross-domain generalization under diverse spoofing systems, source corpora, and lossy compression conditions. Dataset statistics are summarized in Table~\ref{tab:dataset_statistics}. Following the official ASVspoof evaluation protocols, we report the equal error rate (EER) as the performance metric.

\subsection{Implementation Details}
Audio signals are resampled to 16~kHz and fixed to 4 seconds ($64,000$ samples) via sequential repetition or cropping (random during training, center during evaluation). The downstream backbone hidden dimension is $d = 128$, and all sequence operators (MHSA, GDN2, Raven) as well as the MHGAP layer use $K = 4$ heads.

Models are trained for up to 20 epochs with a batch size of 32 and early stopping after 7 epochs. We optimize using AdamW ($\beta_1 = 0.9, \beta_2 = 0.95$, weight decay 0.1) \cite{loshchilov2017decoupled} and Focal Loss \cite{lin2017focal} with class weights $\alpha = [0.8, 0.2]$ (prioritizing class 0, bonafide), with an initial learning rate of $1 \times 10^{-4}$ governed by a cosine annealing schedule with 5\% linear warmup. Training employs BF16/FP32 mixed precision. RawBoost data augmentation (Algo 4) \cite{tak2021rawboost} is applied only in specific ablation setups. All models are trained with fixed random seeds on NVIDIA RTX 4090 and RTX 4080 Super GPUs.

\begin{table}[t]
    \centering
    \small
    \begin{tabular}{llccc}
        \toprule
        SSL frontend
        & Operator
        & 19LA & 21LA & 21DF \\
        \midrule

        \multirow{3}{*}{HuBERT Base}
        & MHSA  & 6.45          & \textbf{8.78}  & 18.27 \\
        & GDN2  & \textbf{5.52} & 9.73           & \textbf{15.15} \\
        & Raven & 7.31          & 12.53          & 15.65 \\
        \midrule

        \multirow{3}{*}{wav2vec 2.0 Base}
        & MHSA  & 4.48          & 12.08          & 19.20 \\
        & GDN2  & 4.44          & 13.84          & 18.49 \\
        & Raven & \textbf{4.33} & \textbf{9.77}  & \textbf{17.72} \\
        \midrule

        \multirow{3}{*}{WavLM Base}
        & MHSA  & \textbf{6.67} & \textbf{11.47} & \textbf{15.58} \\
        & GDN2  & 8.83          & 13.00          & 17.35 \\
        & Raven & 11.24         & 13.86          & 16.70 \\
        \midrule

        \multirow{3}{*}{WavLM Base+}
        & MHSA  & 7.17          & \textbf{14.71} & 17.25 \\
        & GDN2  & \textbf{6.14} & 14.88          & 16.07 \\
        & Raven & 11.27         & 15.94          & \textbf{14.28} \\
        \bottomrule
    \end{tabular}
    \caption{
        EER (\%) results of 6-layer baseline backbones using frozen SSL frontends. Lower is better.
    }
    \label{tab:frozen_ssl_baselines}

\end{table}

\section{Experiments}

\begin{table*}[t]
    \centering
    \small

    \begin{tabular}{cl*{12}{c}}
        \toprule

        \multicolumn{14}{c}{\textbf{Weight-Shared Recurrent Latent Refinement}} \\
        \midrule

        \multirow{3}{*}{Passes}
        & \multirow{3}{*}{Operator}
        & \multicolumn{3}{c}{\textbf{HuBERT Base}}
        & \multicolumn{3}{c}{\textbf{wav2vec 2.0 Base}}
        & \multicolumn{3}{c}{\textbf{WavLM Base}}
        & \multicolumn{3}{c}{\textbf{WavLM Base+}} \\
        \cmidrule(lr){3-5}
        \cmidrule(lr){6-8}
        \cmidrule(lr){9-11}
        \cmidrule(lr){12-14}

        &
        & 19LA & 21LA & 21DF
        & 19LA & 21LA & 21DF
        & 19LA & 21LA & 21DF
        & 19LA & 21LA & 21DF \\

        \midrule

        \multirow{3}{*}{2}
        & MHSA
        & 9.52 & 13.90 & 19.06
        & \textbf{4.56} & \textbf{10.14} & \textbf{20.04}
        & \textbf{9.25} & \textbf{13.81} & \textbf{16.85}
        & \textbf{7.05} & 15.81 & 29.81 \\

        & GDN2
        & \textbf{7.02} & \textbf{11.60} & \textbf{17.61}
        & 5.38 & 11.27 & 21.96
        & 12.12 & 15.44 & 20.49
        & 9.57 & \textbf{15.10} & \textbf{20.24} \\

        & Raven
        & 9.61 & 12.95 & 29.67
        & 12.65 & 16.46 & 25.05
        & 17.89 & 19.51 & 22.79
        & 12.93 & 16.26 & 26.01 \\

        \cmidrule(lr){1-14}

        \multirow{3}{*}{3}
        & MHSA
        & 8.01 & \textbf{11.01} & \textbf{15.97}
        & 7.47 & 12.27 & \textbf{20.62}
        & \textbf{6.86} & \textbf{11.84} & \textbf{17.80}
        & \textbf{8.52} & 17.05 & 23.58 \\

        & GDN2
        & \textbf{7.29} & 12.53 & 19.41
        & \textbf{4.46} & \textbf{9.62} & 21.19
        & 10.27 & 13.74 & 20.16
        & 9.13 & \textbf{13.29} & \textbf{21.71} \\

        & Raven
        & 15.17 & 16.16 & 28.76
        & 13.32 & 23.27 & 30.01
        & 18.98 & 20.96 & 24.74
        & 14.01 & 17.29 & 25.34 \\

        \bottomrule
    \end{tabular}
    \caption{
        Table 3: EER (\%) of weight-shared recurrent backbones with two and three recurrent passes. All SSL frontends are frozen. The best result for each SSL frontend, recurrent setting, and evaluation set is highlighted in bold.
    }
    \label{tab:looped_models}
\end{table*}

\begin{table*}[htbp]
    \centering
    \small

    \begin{tabular}{cl*{12}{c}}
        \toprule
        \multicolumn{14}{c}{\textbf{Homogeneous HRM}} \\
        \midrule
        \multirow{3}{*}{Schedule}
        & \multirow{3}{*}{Operator}
        & \multicolumn{3}{c}{\textbf{HuBERT Base}}
        & \multicolumn{3}{c}{\textbf{wav2vec 2.0 Base}}
        & \multicolumn{3}{c}{\textbf{WavLM Base}}
        & \multicolumn{3}{c}{\textbf{WavLM Base+}} \\
        \cmidrule(lr){3-5}
        \cmidrule(lr){6-8}
        \cmidrule(lr){9-11}
        \cmidrule(lr){12-14}

        & 
        & 19LA & 21LA & 21DF
        & 19LA & 21LA & 21DF
        & 19LA & 21LA & 21DF
        & 19LA & 21LA & 21DF \\

        \midrule

        \multirow{3}{*}{$H_2L_1$}
        & MHSA
        & 11.58 & 13.82 & 25.31
        & 14.30 & 14.25 & 27.42
        & 14.83 & 18.97 & \textbf{21.74}
        & 15.55 & 16.46 & 24.36 \\

        & GDN2
        & \textbf{10.36} & \textbf{11.27} & \textbf{23.27}
        & \textbf{9.57} & \textbf{12.80} & 26.57
        & \textbf{11.97} & \textbf{15.35} & 24.49
        & 14.60 & 17.13 & \textbf{23.88} \\

        & Raven
        & 14.04 & 14.79 & 25.21
        & 13.27 & 15.02 & \textbf{24.72}
        & 13.99 & 17.01 & 24.81
        & \textbf{14.39} & \textbf{16.17} & 27.00 \\

        \cmidrule(lr){1-14}

        \multirow{3}{*}{$H_2L_2$}
        & MHSA
        & 13.02 & 15.18 & 22.85
        & \textbf{8.66} & \textbf{12.09} & \textbf{24.40}
        & \textbf{13.17} & \textbf{17.00} & 29.04
        & 17.85 & 17.20 & 36.71 \\

        & GDN2
        & \textbf{10.24} & \textbf{12.51} & \textbf{20.28}
        & 11.03 & 13.45 & 24.70
        & 13.47 & 17.04 & 27.72
        & \textbf{12.68} & \textbf{15.09} & \textbf{19.92} \\

        & Raven
        & 15.70 & 14.84 & 25.19
        & 13.70 & 18.33 & 26.11
        & 16.50 & 17.64 & \textbf{26.46}
        & 21.50 & 16.87 & 44.70 \\

        \cmidrule(lr){1-14}

        \multirow{3}{*}{$H_2L_3$}
        & MHSA
        & \textbf{9.23} & 11.46 & 24.24
        & 11.48 & 13.53 & \textbf{22.08}
        & \textbf{8.93} & \textbf{13.46} & \textbf{24.09}
        & 15.67 & 16.78 & 23.25 \\

        & GDN2
        & 9.24 & \textbf{10.88} & \textbf{21.16}
        & \textbf{8.47} & \textbf{10.40} & 24.28
        & 12.10 & 14.54 & 24.64
        & \textbf{12.46} & \textbf{15.04} & \textbf{16.97} \\

        & Raven
        & 17.93 & 17.66 & 25.21
        & 15.53 & 15.91 & 25.01
        & 17.49 & 17.64 & 27.31
        & 14.86 & 16.36 & 22.01 \\

        \bottomrule
    \end{tabular}
    \caption{
        EER (\%) of homogeneous hierarchical backbones under different recurrent schedules. All SSL frontends are frozen. The best result for each SSL frontend and evaluation set is highlighted in bold.
    }
    \label{tab:homogeneous_hrm}
\end{table*}

\subsection{Single-Pass Backbones}

Table~\ref{tab:frozen_ssl_baselines} presents single-pass baseline results under frozen SSL frontends. Paired with W2V2B, all backbones achieved low 19LA EERs, with Raven yielding the best 19LA EER of 4.33\% (and 9.77\%/17.72\% on 21LA/21DF). However, Raven's performance degraded when combined with WavLMB or WavLMB+ compared to HuB or W2V2B. These results demonstrate a clear interaction between SSL features and sequence operators: W2V2B excels in-domain (19LA), while the WavLM family shows advantages in cross-domain scenarios.

\subsection{Weight-Shared Recurrent}

To test whether gains stem merely from extra computation, we repeatedly executed the same module for 2 or 3 iterations without adding parameters (Table~\ref{tab:looped_models}). For HuB, increasing MHSA loops from 2 to 3 improved performance across all benchmarks, reducing 21DF EER from 19.06\% to 15.97\% (a 16.21\% relative drop). On W2V2B, 3-loop GDN2 achieved 4.46\% and 9.62\% EERs on 19LA and 21LA, outperforming its 2-loop setting. Conversely, Raven degraded with more loops. These findings indicate that increasing effective depth does not necessarily refine features, as naive repetition can introduce redundancy or error accumulation.

\begin{table*}[t]
    \centering
    \small
    \begin{tabular}{cl*{12}{c}}
        \toprule

        \multicolumn{14}{c}{\textbf{Heterogeneous HRM}} \\
        \midrule

        \multirow{3}{*}{Schedule}
        & \multirow{3}{*}{System}
        & \multicolumn{3}{c}{\textbf{HuBERT Base}}
        & \multicolumn{3}{c}{\textbf{wav2vec 2.0 Base}}
        & \multicolumn{3}{c}{\textbf{WavLM Base}}
        & \multicolumn{3}{c}{\textbf{WavLM Base+}} \\
        \cmidrule(lr){3-5}
        \cmidrule(lr){6-8}
        \cmidrule(lr){9-11}
        \cmidrule(lr){12-14}

        &
        & 19LA & 21LA & 21DF
        & 19LA & 21LA & 21DF
        & 19LA & 21LA & 21DF
        & 19LA & 21LA & 21DF \\

        \midrule

        \multirow{2}{*}{$H_2L_1$}
        & HALG
        & \textbf{12.27} & 15.99 & \textbf{21.88}
        & \textbf{9.66} & \textbf{14.19} & 25.76
        & \textbf{9.93} & \textbf{12.25} & \textbf{21.04}
        & \textbf{13.78} & \textbf{15.28} & \textbf{21.73} \\

        & HALR
        & 13.84 & \textbf{15.81} & 22.20
        & 12.51 & 17.10 & \textbf{25.19}
        & 14.29 & 16.75 & 25.36
        & 16.61 & 17.42 & 26.27 \\

        \cmidrule(lr){1-14}

        \multirow{2}{*}{$H_2L_2$}
        & HALG
        & \textbf{11.40} & \textbf{12.63} & \textbf{20.75}
        & \textbf{7.32} & \textbf{10.89} & \textbf{22.46}
        & 13.99 & 16.56 & 24.20
        & \textbf{14.15} & \textbf{16.08} & \textbf{22.08} \\

        & HALR
        & 13.21 & 15.30 & 29.77
        & 9.67 & 12.60 & 23.10
        & \textbf{11.99} & \textbf{15.21} & \textbf{23.42}
        & 16.74 & 17.16 & 23.79 \\

        \cmidrule(lr){1-14}

        \multirow{2}{*}{$H_2L_3$}
        & HALG
        & \textbf{15.04} & \textbf{16.74} & \textbf{24.30}
        & 10.86 & \textbf{12.74} & 22.32
        & 14.33 & 15.76 & 24.19
        & \textbf{15.13} & \textbf{16.43} & 21.27 \\

        & HALR
        & 20.63 & 20.62 & 31.38
        & \textbf{9.35} & 16.01 & \textbf{19.17}
        & \textbf{12.76} & \textbf{14.50} & \textbf{22.22}
        & 15.75 & 26.59 & \textbf{17.31} \\

        \bottomrule
    \end{tabular}
    \caption{
        EER (\%) of heterogeneous HRM configurations under different recurrent schedules. 
    }
    \label{tab:heterogeneous_hrm}
\end{table*}

\begin{table*}[tb]
    \centering
    \small

    \begin{tabular}{llcccccc}
        \toprule
        \multirow{2}{*}{Architecture}
        & \multirow{2}{*}{Schedule}
        & \multicolumn{2}{c}{19LA}
        & \multicolumn{2}{c}{21LA}
        & \multicolumn{2}{c}{21DF} \\
        \cmidrule(lr){3-4}
        \cmidrule(lr){5-6}
        \cmidrule(lr){7-8}
        &
        & w/o DA & w/ DA
        & w/o DA & w/ DA
        & w/o DA & w/ DA \\
        \midrule

        Baseline
        & 6-layer
        & 1.50 & 4.47
        & 7.26 & 5.14
        & 12.07 & 9.54 \\

        \midrule

        \multirow{2}{*}{Looped}
        & 2 passes
        & 6.09 & 10.94
        & 10.39 & 10.12
        & 21.42 & 14.49 \\

        & 3 passes
        & 6.30 & 12.13
        & 10.93 & 9.39
        & 19.62 & 14.23 \\

        \midrule

        \multirow{3}{*}{Homo-HRM}
        & $H_2L_1$
        & 1.74 & 4.80
        & 8.15 & 5.31
        & 13.53 & 11.04 \\

        & $H_2L_2$
        & \textbf{1.54} & \textbf{4.55}
        & \textbf{6.78} & \textbf{4.89}
        & \textbf{13.34} & 9.94 \\

        & $H_2L_3$
        & 1.56 & 4.65
        & 7.89 & 5.04
        & 13.84 & \textbf{9.59} \\

        \midrule

        \multirow{3}{*}{Hetero-HRM}
        & $H_2L_1$
        & 1.65 & \textbf{4.43}
        & \textbf{5.85} & \textbf{4.70}
        & 14.78 & 10.29 \\

        & $H_2L_2$
        & \textbf{1.36} & 4.52
        & 6.72 & 4.93
        & \textbf{11.64} & 10.69 \\

        & $H_2L_3$
        & 1.87 & 6.21
        & 6.72 & 6.20
        & 12.70 & \textbf{9.69} \\

        \bottomrule
    \end{tabular}
    \caption{
        Ablation results under selective SSL fine-tuning and different data augmentation settings (reported in EER (\%)).
    }
    \label{tab:top2_da_ablation}
\end{table*}

\subsection{Homogeneous HRM}

Table~\ref{tab:homogeneous_hrm} compares homogeneous HRM configurations. On HuB, $H_2L_3$ MHSA and GDN2 achieved 9.23\% and 10.88\% EERs on 19LA and 21LA, respectively, failing to surpass single-pass baselines. On W2V2B, $H_2L_3$-GDN2 achieved the best homogeneous results, yet remained weaker than certain single-pass or Looped baselines. WavLM variants were generally uncompetitive. Compared to standard models, homogeneous HRM showed more performance fluctuations; increasing low-level updates brought no consistent gains, and optimal schedules varied across frontends and test conditions.

\subsection{Heterogeneous HRM}
Homogeneous experiments indicate that decoupling update frequencies alone is insufficient to consistently improve detection performance. Based on this observation, we further assign MHSA to the high-level module while adopting GDN2 and Raven in the low-level module, forming two heterogeneous configurations: HALG and HALR. Table~\ref{tab:heterogeneous_hrm} presents the corresponding results. The heterogeneous design markedly outperformed homogeneous HRM under several frontend and schedule settings: for example, W2V2B-$H_2L_2$-HALG achieved EERs of 7.32\% and 10.89\% on 19LA and 21LA, outperforming most homogeneous counterparts under the same schedule; WavLMB-$H_2L_1$-HALG yielded relatively balanced EERs of 9.93\%, 12.25\%, and 21.04\% across the three evaluation sets. In most setups, HALG performed better than HALR. These results demonstrate that while complementary sequence operators can alleviate redundancy from homogeneous H/L updates, their gains may depend on whether frontend representations can adapt to downstream hierarchical updates. Guided by this insight, we unfreeze the top two Transformer layers of the SSL encoder in the final ablation study.

\subsection{Ablation Study of GDN2}

Based on the overall performance in the frozen-frontend experiments, we selected W2V2B and GDN2 for the final ablation study, unfreezing the top two Transformer layers of the SSL encoder. Table~\ref{tab:top2_da_ablation} presents the ablation results across different configurations. Without data augmentation, selective fine-tuning significantly reduced the EERs of all major models: Hetero-$H_2L_2$ achieved EERs of 1.36\%, 6.72\%, and 11.64\% on 19LA, 21LA, and 21DF, consistently outperforming the GDN2 baseline; Hetero-$H_2L_1$ achieved the lowest 21LA EER of 5.85\%. This indicates that when high-level SSL representations participate in task adaptation, heterogeneous high/low-level configurations can obtain performance advantages over matched baselines across diverse evaluation conditions. In contrast, the overall performance of weight-shared recurrent models remained inferior, suggesting that such structures are less suitable as detection backbones under current settings. Data augmentation significantly improved cross-domain performance, though at the cost of reduced accuracy on 19LA: for instance, the 19LA EER of Hetero-$H_2L_1$ increased from 1.65\% to 4.43\%, whereas its 21LA and 21DF EERs dropped from 5.85\% and 14.78\% to 4.70\% and 10.29\% (relative reductions of 19.66\% and 30.38\%), respectively. These results demonstrate that data augmentation enhances model robustness and cross-domain generalizability.Notably, the downstream backbone parameter counts for the GDN2 baseline and Hetero-HRM are 1.405 M and 1.252 M, respectively. Hetero-HRM reduces backbone parameters by 10.89\% while achieving competitive performance under several evaluation conditions, highlighting its strong competitiveness for the SDD task.

\section{Conclusion}
This paper presents REIMU, a controlled investigation of recurrent hierarchical architectures for SSL-based speech deepfake detection. By comparing single-pass backbones, weight-shared recurrence, homogeneous HRM, and heterogeneous HRM, we disentangle the effects of recurrent computation, parameter sharing, and operator assignment. The results show that recurrence or hierarchical decomposition alone does not guarantee improved detection, whereas complementary high- and low-level operators provide a more competitive design. Moreover, the heterogeneous configurations remain competitive with fewer downstream parameters, demonstrating the potential of parameter-efficient hierarchical reasoning for SDD.




\bibliography{aaai2027}


\end{document}